\documentclass[aps,twocolumn,showpacs,superscriptaddress]{revtex4-1}
\usepackage{graphicx,mathtools}
\usepackage[dvipdfmx,colorlinks,linkcolor=blue,anchorcolor=blue,citecolor=blue,urlcolor=blue]{hyperref}

\def\avg#1{\langle#1\rangle}

\def\be{\begin{equation}} \def\ee{\end{equation}}
\def\bea{\begin{eqnarray}} \def\eea{\end{eqnarray}}

\def\nn{\nonumber}

\begin{document}
\title{Competing orders in the 2D half-filled SU($2N$) Hubbard model through
the pinning field quantum Monte-Carlo simulations}
\author{Da Wang}\email{d6wang@physics.ucsd.edu}
\affiliation{Department of Physics, University of California, 
San Diego, California 92093, USA}
\author{Yi Li}
\affiliation{Department of Physics, University of California, 
San Diego, California 92093, USA}
\affiliation{
Princeton Center for Theoretical Science, Princeton University, Princeton, New Jersey 08544, USA}
\author{Zi Cai}
\affiliation{
Department of Physics and Arnold Sommerfeld Center for Theoretical Physics,
Ludwig-Maximilians-Universit{\"a}t M{\"u}nchen, 
Theresienstra{\ss}e 37, 80333 Munich, Germany}
\author{Zhichao Zhou}
\affiliation{
School of Physics and Technology, Wuhan University, Wuhan 430072, China}
\author{Yu Wang}\email{yu.wang@whu.edu.cn}
\affiliation{
School of Physics and Technology, Wuhan University, Wuhan 430072, China}
\author{Congjun Wu}\email{wucj@physics.ucsd.edu}
\affiliation{Department of Physics, University of California, 
San Diego, California 92093, USA}

\begin{abstract}
We non-perturbatively investigate the ground state magnetic properties of 
the 2D half-filled SU($2N$) Hubbard model in the square lattice by using 
the projector determinant quantum Monte Carlo simulations combined with 
the method of local pinning fields.
Long-range Neel orders are found for both the SU(4) and SU(6) cases at 
small and intermediate values of $U$.
In both cases, the long-range Neel moments exhibit non-monotonic behavior 
with respect to $U$, which first grow and then drop as $U$ increases.
This result is fundamentally different from the SU(2) case in which the 
Neel moments increase monotonically and saturate.
In the SU(6) case, a transition to the columnar dimer phase is found in 
the strong interaction regime.
\end{abstract}
\maketitle

The ultra-cold atom systems have opened up a wonderful opportunity for 
studying novel phenomena which are not easily accessible in usual 
solid state systems. 
For example, the large-spin ultra-cold alkali and alkaline-earth fermions 
exhibit quantum magnetic properties fundamentally different from 
the large-spin solid state systems such as transition metal oxides
\cite{Wu2012}.
In solids, Hund's rule coupling combines several electrons on the same 
cation site into states carrying large spin $S$.
However, the symmetry of these systems is usually only SU(2).
The leading order coupling between two neighboring sites 
is mediated by exchanging one pair of electrons no matter how large
$S$ is, thus quantum spin fluctuations are suppressed by 
the $1/S$-effect.
In contrast, large-hyperfine-spin ultra-cold fermion systems can
possess high symmetries of SU($2N$) and Sp($2N$). 
For the simplest case of spin-$\frac{3}{2}$, a generic Sp(4) symmetry
was proved without fine-tuning, which includes the SU(4) symmetry
as a special case \cite{Wu2003,*Wu2006}.
Such a high symmetry gives rise to exotic properties in 
quantum magnetism and pairing superfluidity 
\cite{Wu2005,Hattori2005,Lecheminant2005,Controzzi2006,Cazalilla2009,
WU2010,Rodriguez2010,Corboz2011,Hung2011,Szirmai2011a}.
Furthermore, large-spin alkaline-earth fermion systems have been 
experimentally realized in recent years \cite{DeSalvo2010,
Taie2010,Krauser2012}.
In particular, an SU(6) Mott insulator of $^{173}$Yb has also been 
observed \cite{Taie2012,Wu2012}.
The above theoretical and experimental progress has stimulated a great deal
of interests in exploring novel properties of
strongly correlated systems with high symmetries
\cite{Hermele2009,Gorshkov2010,Cai2013,Messio2012,Cai2013a,Szirmai2011,
Sinkovicz2013}.

The SU($2N$) Heisenberg model was first introduced into
condensed matter physics to apply the large-$N$ technique to 
systematically handle strong correlation effects in the context of
high $T_c$ cuprates \cite{Affleck1985,
Arovas1988, Affleck1988,Read1989a,Read1989}. 
It was found that on 2D bipartite lattices the SU(2) 
Heisenberg model displays long-range Neel ordering \cite{Anderson1952}.
As $2N$ increases, enhanced quantum fluctuations suppress Neel ordering
and the ground states eventually become dimerized
\cite{Read1989a,Read1989}. 
This transition has been observed by quantum Monte Carlo (QMC) simulations 
\cite{Harada2003,Kawashima2007, Beach2009,Kaul2012,Paramekanti2007,Assaad2005}
for certain representations of the SU($2N$)symmetry
\footnote{The representations of SU($2N$) 
are classified by the Young tableau. 
On a bipartite lattice, the two sublattices can realize two different 
representations of the SU($2N$) group which are complex conjugates 
to each other, such that two neighboring sites can form
an SU($2N$) invariant singlet.
The system can be simply thought 
as loading $m$ fermions per site in the $A$-sublattice  and $2N-m$
fermions per site in the $B$-sublattice.
The case of $m=1$ was investigated in Refs.
\cite{Harada2003,Kawashima2007, Beach2009,Kaul2012};
while the case of $m=N$, which forms the self-conjugate
representation, was studied in Refs. 
\cite{Paramekanti2007,Assaad2005}.
}.
However, for the self-conjugate representations,
a consensus has not been achieved yet. 
A variational Monte Carlo study \cite{Paramekanti2007} 
found Neel ordering when $2N=2$ and $4$, and 
columnar dimer ordering for $2N\geq6$. However,  
in a determinant QMC calculation \cite{Assaad2005}, 
dimer ordering was found 
at $2N\geq 6$ in agreement with the variational QMC study, 
while for the SU(4) case, neither Neel nor dimer ordering exists
in the Heisenberg limit.

The above Heisenberg-type models neglect charge fluctuations.
The interplay between charge and spin degrees of freedom is contained 
in the SU($2N$) Hubbard model \cite{Lu1994,Honerkamp2004,Cai2013a}.
However, owing to the lack of non-perturbative methods, 
the SU($2N$) Hubbard model receives much less attention.
To the best of our knowledge, a systematic non-perturbative
study of the ground state properties of the 2D half-filled models 
is still missing. 
It is even not clear whether Neel or dimer ordering exists 
in the weak, intermediate and strong coupling regimes, respectively.

In this article, we perform a non-perturbative determinant 
QMC study on the half-filled SU($2N$) Hubbard model in the 2D square lattice. 
The ground state magnetic properties are investigated 
by using the local pinning field method 
which directly measures the spatial decay
of the induced order parameters \cite{White2007}. 
Long-range Neel order is identified at weak and intermediate
values of $U$ in the SU($2N$) Hubbard models of
$2 \le 2N \le 6$ we studied.
In the cases of SU(4) and SU(6), the Neel moments first grow then drop
with increasing $U$. 
Furthermore, a transition from the Neel-ordering phase into the columnar 
dimer-ordering phase is observed at a large value of $U$ in the SU(6) case.
This transition is conceivably owing to the competition between 
the weak coupling physics of Fermi surface nesting and strong 
coupling local moment physics.

%%%%%%%%%%%%%%%%%%%%%%%%%%%%%%%%%%%%%%%%%%%%%%%%%%%%%%%%%%%%%%%%
We consider the SU($2N$) Hubbard model in the 2D %$L\times L$ 
square lattice 
with the periodic boundary condition as,
\bea
H=-t\sum_{\langle i,j\rangle,\alpha}\left(c_{i\alpha}^\dag c_{j\alpha}
+h.c.\right)+\frac{U}{2}\sum_{i}\left(n_i-N\right)^2,
\label{eq:hamilton}
\eea
where $t$ is the nearest neighbor hopping integral ($t=1$ in the below);
$U$ is the on-site repulsion; $\alpha$ is the spin index running from $1$ 
to $2N$; $n_i=\sum_{\alpha=1}^{2N} n_{i\alpha}$is the total fermion number 
operator on site $i$.
Eq. \ref{eq:hamilton} possesses the particle-hole symmetry $c_{i\alpha}
\rightarrow (-)^i c_{i\alpha}^\dagger$, which means that it is at half-filling.
In this case, it is well-known that Eq. \ref{eq:hamilton} is free of the 
sign problem for all the values of $N$.

We employ the projector QMC to investigate its quantum magnetic properties
in the ground states.
In QMC studies, the long-range ordering is usually obtained through 
the finite-size scaling of the corresponding structural factors, or, 
correlation functions.
Assuming that the system size is $L\times L$, 
the extrapolated values as $L\rightarrow\infty$ are 
proportional to the magnitude square of order parameters.
Thus it is difficult to distinguish the weakly ordered states
from the truly disordered ones. 
For this reason, there has been a debate whether a quantum spin liquid 
phase exists near the Mott transition in the honeycomb lattice
\cite{Meng2010,Li2011,Sorella2012,Hassan2013,Assaad2013a,*Assaad2012}. 
To overcome this difficulty, we use the pinning field method
\cite{White2007,Assaad2013a,*Assaad2012}, and 
measure the spatial decay of the induced order parameters.
Order parameters instead of their magnitude square are measured, 
and thus numerically they are more sensitive to weak orderings.
This method has also been used in the projector QMC recently
\cite{Assaad2013a}.
To decouple the interaction term, we adopt the Hubbard-Stratonovich (HS)
transformation in the density channel which involves complex numbers
\cite{Hirsch1983}. 
We have designed a new discrete HS decomposition which is exact for 
the cases from SU(2) to SU(6) Hubbard models, and the
algorithm details can be found in the Supplementary Material. 
\footnote{See Supplementary Material [url], which includes Refs.
\cite{Assaad2008,Wu2005a,Schulz1990}. }
Unless specifically stated, the following
parameters are used in simulations: the projection time $\beta=240$ 
and the discretized imaginary time step $\Delta\tau=0.05$. 

Next we use the pinning field method to study the magnetic long-range 
order of the SU($2N$) Hubbard model. 
We define the SU($2N$) generators as
$S^{\alpha\beta}_i=c^\dagger_{i,\alpha}c_{i,\beta}-
\frac{\delta^{\alpha\beta}}{2N} n_i$.
At half-filling, in the Heisenberg limit in which 
charge fluctuations are neglected, each site belongs to the
self-conjugate representation with one column of $N$ boxes.
Without loss of generality, the classic Neel state
configuration can be chosen as follows: each site in 
sublattice $A$ is filled with $N$ fermions from components 
$1$ to $N$, while that in sublattice $B$ is filled 
with components from $N+1$ to $2N$. 
We define the magnetic moment operator on each site $i$ as
\bea
m_i= \frac{1}{2N}
\big\{\sum_{\alpha=1}^{N} S^{\alpha\alpha}_i-\sum_{\alpha=N+1}^{2N} S^{\alpha\alpha}_i
\big\}.
\label{eq:neel}
\eea 
For the configuration defined above, the value of the classic Neel moment
is $m_i=(-)^i\frac{1}{2}$. 
Within the zero temperature projector QMC method, good quantum 
numbers are conserved during the projection.
Thus we use a pair of pinning fields on two neighboring sites 
with a Neel configuration to maintain the
relation $\avg{G|\sum_i S^{\alpha\alpha}_i|G}=0$ for every $\alpha$.
The pinning field Hamiltonian is
\bea
H_{pin,n}= 2Nh_{i_0j_0} \big\{ m_{i_0} - m_{j_0} \big\},
\label{eq:pinning}
\eea
where $i_0$ and $j_0$ are two neighboring sites
defined as $i_0=(1,1)$ and $j_0=(2,1)$, 
respectively.
The initial trial wavefunctions can be chosen as the half-filled
plane-wave states.
The Hamiltonian Eq. \ref{eq:hamilton} plus Eq. \ref{eq:pinning} remains 
free of the sign problem at half-filling.

%---------------------------------------------------------------
\begin{figure}
\includegraphics[width=0.8\linewidth,height=0.55\linewidth]{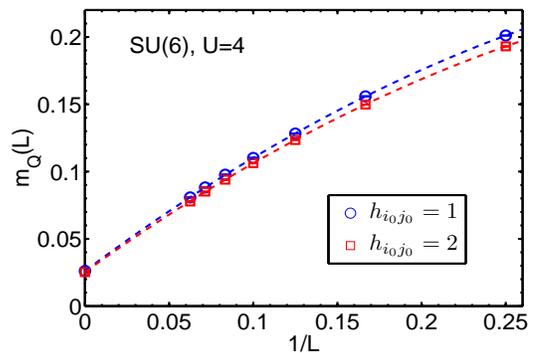}
\caption{Finite size scaling of the residual Neel moment $m_Q(L)$ v.s. 
$1/L$ under pinning fields described by Eq. \ref{eq:pinning} with 
$h_{i_0j_0}=1$ and $2$. 
The largest value of $L$ is 16.
The quadratic polynomial fitting is used. Error bars are smaller than symbols. 
}
\label{fig:diffh}
\end{figure}
%---------------------------------------------------------------

Because the pinning fields in Eq. \ref{eq:neel} break the $SU(2N)$ 
symmetry, the induced magnetic moments prefer the direction defined 
in Eq. \ref{eq:neel}.
The distribution of $m_i$ is staggered with decaying magnitudes
as away from two pinned sites $i_0$ and $j_0$. 
The Neel order parameter is its Fourier component at the wavevector
$Q=(\pi,\pi)$ defined as $m_Q(L)=\frac{1}{L^2}\sum_i (-)^i m_i$.
The long-range order $m_Q$ can be extrapolated as the limit of
\bea
m_Q=\lim_{L\rightarrow \infty} m_Q(L).
\eea
This is because the Fourier component of the pinning field at $Q$ 
is $h_Q=2h_{i_0j_0}/L^2$, which goes to zero as $L\rightarrow \infty$ 
for any finite value of $h_{i_0j_0}$.

To illustrate the sensitivity of the pinning field method to weak
orders, we present the simulations for the SU(6) case
of Eq. \ref{eq:hamilton} with $U=4$.
The finite size scalings of $m_Q(L)$ are presented in Fig. \ref{fig:diffh}
for two different values of $h_{i_0j_0}=1$ and $2$. 
Their extrapolated values as $1/L \rightarrow 0$ are 
$0.0261\pm0.0008$ and $0.0253\pm0.0009$, respectively, which are consistent
with each other and confirm the validity of this method.
Such a small moment is hard to identify using the finite
size scaling of the structural factors, as shown in the
Supplementary Material
and related works\cite{Meng2010,Sorella2012,Assaad2013a,*Assaad2012}. 
Another observation is that the induced values of $m_Q(L)$ are weaker at 
$h_{i_0j_0}=2$ than those at $h_{i_0j_0}=1$ at finite values of $L$,
which shows non-linear correlations between the pinning
centers and the measured sites.
Certainly they converge in the limit of $1/L\rightarrow 0$. 
In the following, we only present the results of $h_{i_0j_0}=2$. 

One may question whether the pinning field method overestimates the tendency
of long-range ordering.
In the Supplementary Material, we apply it to the 1D SU(2) and SU(4) 
Hubbard chains at half-filling.
In the SU(2) case, the ground state is known as a gapless spin liquid, while in 
the SU(4) case, it is gapped with dimerization.
The pinning field method shows the absence of long-range Neel ordering 
in both cases and the asymptotic behavior of power-law spin correlations in 
the case of SU(2).
This further confirms the validity of this method.

%--------------------------------------------------------------
\begin{figure}
\includegraphics[width=\linewidth]{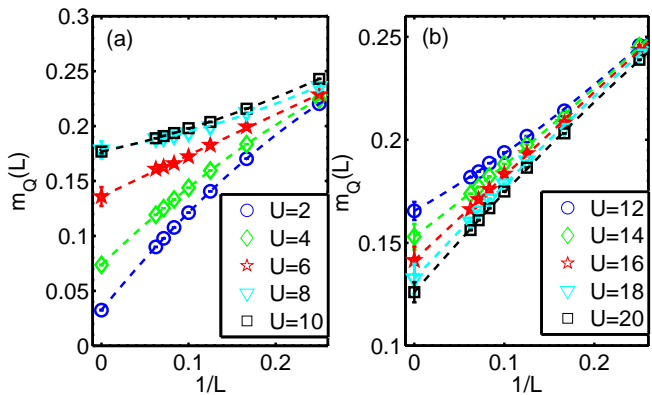}
\caption{Finite size scalings of $m_Q(L)$ v.s. $1/L$ for the 
half-filled SU(4) Hubbard model with different values of $U$.
The largest size is $L=16$.
The quadratic polynomial fitting is used.
Error bars of QMC data are smaller than symbols. 
}
\label{fig:neel_su4}
\end{figure}
%-------------------------------------------------------

We further test the validity of the pinning field method in the 
extensively studied half-filled SU(2) Hubbard model in the square lattice
by QMC \cite{Hirsch1985,Varney2009}.
The long-range Neel ordering we obtained based on 
the pinning field method is consistent with that in
previous QMC literature based on the finite-size scaling
of structure factors. 
Our results are shown in the Supplementary material.
The long-range Neel ordering appears from weak to strong interactions.
The extrapolated values of $m_Q$ increase as $U$ goes up, 
and begin to saturate around $U=10$. 
At $U=20$, $m_Q=0.297\pm0.002$, which is in a good agreement 
with the long-range Neel moment $0.3070(3)$ of the SU(2) Heisenberg 
model \cite{Sandvik1997}.
This behavior is well-known \cite{Hirsch1985,Varney2009}: as $U$ goes up, 
charge fluctuations are suppressed, and thus the low energy physics 
is described by the Heisenberg model.

Next we simulate the SU(4) Hubbard model and the magnetic
ordering is presented in Fig. \ref{fig:neel_su4}.
Similarly to the SU(2) case, long-range Neel ordering appears
for all the values of $U\le 20$.
At each value of $U$, the extrapolated long-range Neel moment $m_Q$ 
is weaker than that in the SU(2) case, which is a result
of the enhanced quantum fluctuations.
Moreover, a striking new feature appears that the relation
$m_Q$ v.s. $U$ becomes non-monotonic as shown in Fig. 
\ref{fig:neel_su2n} below.
The Neel moment $m_Q$ reaches the maximum around $0.178\pm0.008 $ at 
$U \approx 8$, and then decreases as $U$ further increases. 
It remains finite with the largest value of $U=20$ in our simulations.
It is not clear whether $m_Q$ is suppressed
to zero or not in the limit of $U\rightarrow \infty$. 
A previous QMC simulation on the SU(4) Heisenberg model 
shows algebraic spin correlations \cite{Assaad2005}.
It would be interesting to further investigate whether 
the algebraic spin liquid state survives at finite values of $U$.

%-----------------------------------------------------
\begin{figure}%[htbp]
\includegraphics[width=\linewidth]{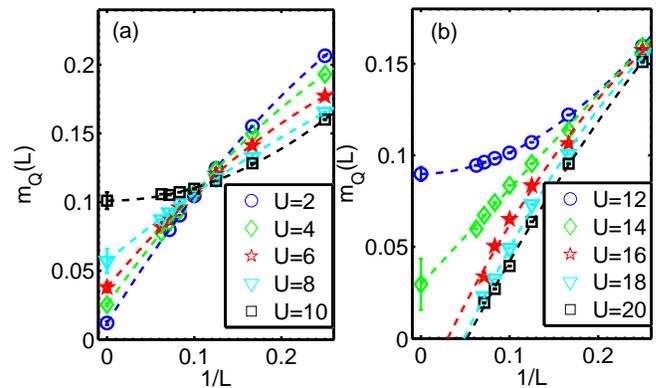}
\caption{Finite size scalings of $m_Q(L)$ v.s. $1/L$ for the SU(6)
Hubbard model at different values of $U$. 
The largest size is $L=16$.
The quadratic polynomial fitting is used.
Error bars of QMC data are smaller than symbols. }
\label{fig:neel_su6}
\end{figure}
%------------------------------------------------------------

With further increases in $2N$, the Neel ordering is more strongly 
suppressed by quantum spin fluctuations.
The finite-size scalings for the SU(6) case at different values of $U$ 
are presented in Fig. \ref{fig:neel_su6}.
For all the values of $U\le 14$, we find nonzero Neel ordering 
by using the quadratic polynomial fitting. 
The extrapolated Neel moment $m_Q$ {\it v.s.} 
$U$ for the SU(6) case are plotted in Fig. \ref{fig:neel_su2n}.
For comparison, those of the SU(2) and SU(4) are also plotted together.
Similar to the SU(4) case, the long-range Neel moments are non-monotonic 
which reach the maximum around $U\approx 10$.
Strikingly, the Neel ordering disappears beyond a critical value of 
$U_c$ which is estimated as $14<U_c<16$.

%%%%%%%%%%%%%%%%%%%%%%%%%%%%%%%%%%%%%%%%%%%%%%%%%%%%%%%%%%%%%%%%%
\begin{figure}
\includegraphics[width=0.9\linewidth]{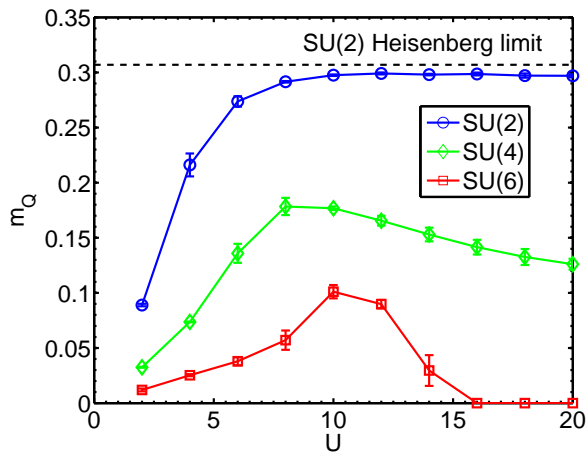}
\caption{The ground state Neel ordering of the 2D half-filled SU($2N$) 
Hubbard model in the square lattice.
The relations of long-range Neel moments $m_Q$ v.s. $U$ 
are plotted for $2N=2,4$ and $6$. For comparison, the SU(2) Heisenberg 
limit result is plotted as the dotted line. 
The error bars are obtained from the least square fittings 
with $95\%$ confidence bounds.}
\label{fig:neel_su2n}
\end{figure}
%%%%%%%%%%%%%%%%%%%%%%%%%%%%%%%%%%%%%%%%%%%%%%%%%%%%%%%%%%%%%%%%%%%%%%%%

\begin{figure}
\includegraphics[width=\linewidth]{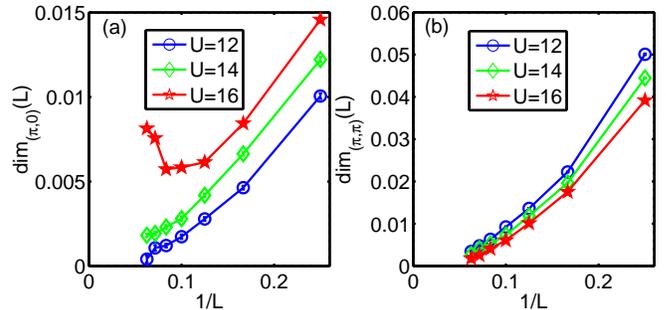}
\caption{Finite size scalings of the dimer order parameters 
in the half-filled SU(6) Hubbard model.
(a) $\mbox{dim}_{(\pi,0)}(L)$ and (b) $\mbox{dim}_{(\pi,\pi)}(L)$
at wavevectors  $Q^\prime=(\pi,0)$ and $Q=(\pi,\pi)$, respectively.
The largest size is $L=16$. Error bars of QMC data are smaller than symbols. 
}
\label{fig:dimer}
\end{figure}

The low energy effective model of half-filled Hubbard models
in the strong coupling regime is the Heisenberg model. 
According to the large-$N$ study of the SU($2N$) Heisenberg model 
with the self-conjugate $1^N$ representation \cite{Read1989a,
Read1989}, dimerization appears in the large-$N$ limit. 
Thus the suppression of the Neel order at large values 
of $U$ is expected from the competing dimer ordering.
To investigate this competition,
we further apply the pinning field method to study the dimer ordering for 
the SU(6) Hubbard model and results are presented in 
Fig. \ref{fig:dimer}.
The following dimer pinning field is applied, which changes the hopping 
integral of a bond $i_0j_0$
\footnote{The kinetic energy dimerization, \textit{i.e.}, the staggered 
ordering of bonding strength, is equivalent to spin-dimerization 
in the large-$U$ limit. In the background of half-filled Mott insulating
states, the kinetic energy on each bond vanishes at the 1st order 
perturbation theory. Its effect begins to appear at the 2nd order as
the antiferromagnetic spin-spin coupling.},
\bea
H_{pin,dim}=-\Delta t_{i_0j_0} \sum_\alpha \Big\{ c_{i_0,\alpha}^\dag 
c_{j_0,\alpha}+h.c. \Big\},
\label{eq:dimer}
\eea
where $i_0$ and $j_0$ are defined before.
The bonding strength between sites $i$ and $i+\hat x$ is defined as 
$d_{i,x}=\frac{1}{2}\langle G| c_{i\alpha}^\dag c_{i+x,\alpha}+h.c. |G\rangle$, where $|G\rangle$ is the ground state.
%A typical spatial distribution of $d_{i,x}$ with the pinning field is 
%plotted in the inset of Fig. \ref{fig:dimer} (b), which shows a staggered 
%configuration along $x$-direction with decaying magnitudes. 
We define the dimer order parameter at the wavevector
$(\pi,0)$ as
\bea
\mbox{dim}_{(\pi,0)}(L)=\frac{1}{L^2} \sum_i (-)^{i_x} d_{i,x},
\eea 
where $i_x$ is the $x$-coordinate of site $i$.
Following the same reasoning to extrapolate the long-range Neel ordering 
before, we define the long-range dimer order parameter
as $\mbox{dim}_{(\pi,0)}=\lim_{L\rightarrow \infty} \mbox{dim}_Q(L)$.
The finite size scalings for $\mbox{dim}_{(\pi,0)}(L)$ are plotted in
Fig. \ref{fig:dimer} (a), which shows the columnar dimerization
appears when $U$ is above a critical value $U_c^\prime$
which is also estimated around $14\sim16$.
It lies in the same interaction regime that Neel ordering starts
to vanish.
However, whether this transition is of second order such 
that $U_c=U_c^\prime$, or, it is of first order, still needs further 
numeric investigation.
We also measure the dimerization at $Q=(\pi,\pi)$
induced by the pinning field Eq. \ref{eq:dimer},
defined as $\mbox{dim}_{(\pi,\pi)}(L) =\frac{1}{L^2}\sum_i (-)^i d_{i,x}$,
whose finite size scaling shows the absence of long-range order.

%%%%%%%%%%%%%%%%%%%%%%%%%%%%%%%%%%%%%%%%%%%%%%%%%%%%%%%%%
%%%%%%%%%%%%%%%%%%%%%%%%%%%%%%%%%%%%%%%%%%%%%%%%%%%%%%%%%%%%%%%%

The nature of the transition between the Neel and dimer orderings
is an interesting question.
In the literature \cite{Senthil2004,Sandvik2007},
ring exchange terms are added to the SU(2) Heisenberg model,
which suppress Neel ordering and lead to dimerization.
However, our SU(6) case is dramatically different. 
The SU(6) Neel ordering appears in the regime of weak and intermediate
interactions.
In this regime ring exchanges are prominent because
they reflect short-range charge fluctuations.
Our results agree with the picture of Fermi surface nesting because
the Neel ordering wavevector $Q=(\pi,\pi)$ is commensurate with the 
Fermi surface at half-filling, while dimerization is not favored
because its wavevector $Q^\prime=(\pi,0)$ does not satisfy 
the nesting condition
\footnote{Even though the $Q=(\pi,\pi)$ nesting 
vector allows for commensurate dimerization, there appears a vertex 
function $f_d(k)=\sin k_x $ in the expression of its susceptibility 
$\chi_{dim}(Q)=-\mbox{Tr}\left(G(k)f_d(k)G(k+Q)f_d(k+Q)\right)$,
where $G(k)$ is the free Green's function. The susceptibility 
for the Neel ordering shares the same expression by substituting
the vertex function with $f_N(k)=1$. The low energy density 
of states concentrate around points of van Hove singularity 
located at $(\pm\pi,0)$ and $(0,\pm \pi)$ at which the vertex 
function $f_d(k)$ vanishes.  Thus the angular dependences
of vertex functions suppress the dimer ordering but favor
Neel ordering at $Q=(\pi,\pi)$ in the weak coupling regime.
}.
On the other hand, local moment physics dominates when deeply inside
the Mott insulating phase in the strong coupling regime.
The exchange energy per site in the dimerized phase is estimated
at the order of $N^2 J$ with $J=4t^2/U$, while that of the Neel state is 
$z N J$ where $z$ is the coordination number. 
Thus dimerization wins when both conditions of large-$U$ and 
large-$N$ limits are met in agreement with previous
theoretical results on SU($2N$) Heisenberg models \cite{Read1989a}.

\textit{Summary.}---
We have applied the method of local pinning fields
in QMC simulations to investigate quantum magnetic properties
of the 2D half-filled SU($2N$) Hubbard model in the square lattice.
This method is sensitive to weak long-range orders.
Long-range Neel ordering is found for the SU(4) case from weak
to strong interactions. 
For the SU(6) case, a transition from the staggered Neel ordering 
to the columnar dimerization is found with increasing $U$. 
The conceivable mechanism is the competition between the weak coupling 
Fermi surface nesting physics and the strong coupling local moment physics. 
The above QMC simulations may provide a reference point for further 
investigating the even more challenging problem of doped SU($2N$) 
Mott-insulators.

\textit{Acknowledgment.}---
We thank J. E. Hirsch, Y. Wan for helpful discussions. 
Especially, we thank H. H. Hung for providing numeric results from
exact diagonalizations for comparison.
D. W., Y. L., and C. W. are supported by the NSF DMR-1105945 and 
AFOSR FA9550-11-1- 0067(YIP); 
Z. C. thanks the German Research Foundation through DFG FOR 801. 
Z. Z., Y. W. and C. W. acknowledge the financial support from the National 
Natural Science Foundation of China (11328403, J1210061), 
and the Fundamental Research Funds for the Central Universities.
Y. L. thanks the Inamori Fellowship and the support at
the Princeton Center for Theoretical Science.
We acknowledge support from the Center for Scientific Computing from the CNSI, MRL: an NSF MRSEC (DMR-1121053) and NSF CNS-0960316.

\bibliography{sun}

%%%%%%%%%%%%%%%%%%%%%%%%%%%%%%%%%%%%%%%%%%%%%%%%%%%%%%%%%%%%%%%%%%%
\newpage
\appendix

\begin{center}
{\Large{Supplementary Material}}
\end{center}

%%%%%%%%%%%%%%%%%%%%%%%%%%%%%%%%%%%%%%%%%%%%%%%%%%%%%%%%%%%%%

In this supplementary material, we explain the algorithm of
the projector quantum Monte Carlo method in Sect. \ref{sect:QMC}.
Various tests of the local pinning field method are presented
in Sect. \ref{sect:test}.
The error analysis is performed in Sect. \ref{sect:error}.

%%%%%%%%%%%%%%%%%%%%%%%%%%%%%%%%%%%%%%%%%%%%%%%%%%%%%%%%%%%%%

\section{Projector quantum Monte Carlo and
Hubbard-Stratonovich decomposition}
\label{sect:QMC}

We adopt the projector determinant QMC method \cite{Assaad2008} to study 
the half-filled SU($2N$) Hubbard model.
The basic idea is to apply
the  projection operator $\mathrm{e}^{-\beta H/2}$
on a trial wave function $|\Psi_T\rangle$.
If $\avg{\Psi_G|\Psi_T}\neq 0$ and there exists
a nonzero gap between $|\Psi_G\rangle$
and the first excited state, $|\Psi_G\rangle$ is arrived
as the projection time $\beta\rightarrow \infty$,
\bea
|\Psi_G\rangle=\lim_{\beta\rightarrow\infty}\mathrm{e}^{-\beta H/2}|\Psi_T\rangle.
\eea
The projection time $\beta$ can be divided into $M$ slices
with $\beta=M\Delta\tau$.

\begin{figure}%[htbp]
\includegraphics[width=0.45\textwidth]{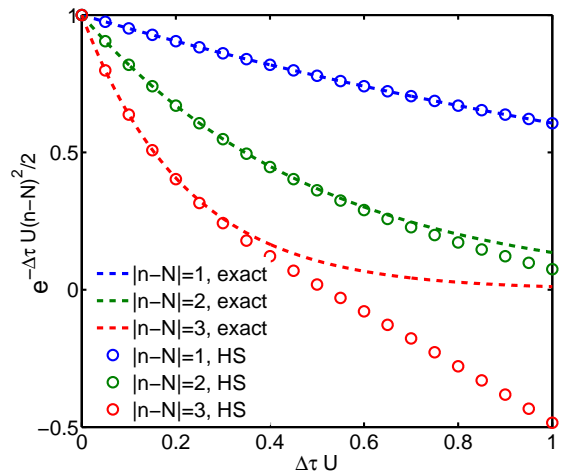}
\caption{Error due to the HS decomposition using parameters
defined in Eq.~\ref{eq:oldHS}. The dashed lines are exact results of
$\mathrm{e}^{-\Delta\tau U (n_i-N)^2/2}$ with $|n-N|=1,2,3$
respectively. The circles represent the results after the HS
transformation.}
\label{fig:HSerror}
\end{figure}

The second order Suzuki-Trotter decomposition is used to separate
the kinetic and interaction energy parts in each time slice,
\bea\label{eq:trotter}
\mathrm{e}^{-\Delta\tau(K+V)}=\mathrm{e}^{-\Delta\tau K/2}
\mathrm{e}^{-\Delta\tau V}\mathrm{e}^{-\Delta\tau K/2}+o[(\Delta\tau)^3],
\eea
where $K$ and $V$ represent the kinetic and interaction terms, respectively.
For the $V$ term, a discrete Hubbard-Stratonovich
(HS) transformation is defined as \cite{Hirsch1983}
\bea
e^{-\lambda^2(n_i-N)^2}=\frac{1}{4} \sum_{\mathclap{l=\pm1,\pm2}}\gamma_i(l)
e^{i\eta_i(l) (n_i-N)}+o[(\Delta\tau)^4], 
\label{eq:approxHS}
\eea 
where $n_i=\sum_{\alpha=1}^{2N} c^\dag_{i\alpha}c_{i\alpha}$;
$\lambda=\sqrt{\Delta\tau U/2}$;
$\gamma$'s and $\eta$'s are 
discrete HS fields given by the following values 
\cite{Assaad1998}
\bea 
\gamma(\pm 1)&=&1+\frac{\sqrt{6}}3, \ \ \, \ \ \, 
\gamma(\pm 2)=1-\frac{\sqrt{6}}3, \nn \\
\eta(\pm 1)&=&\pm\sqrt{\Delta\tau U}\sqrt{3-\sqrt{6}},  \nn \\
\eta(\pm 2)&=&\pm\sqrt{\Delta \tau U} \sqrt{3+\sqrt{6}}. 
\label{eq:oldHS}
\eea 
This decomposition is widely used
in QMC simulations \cite{Wu2005a, Assaad1998}. 
However, one should be careful that at large values of $U$ and $|n-N|$
in Eq. \ref{eq:approxHS}. 
In Fig.~\ref{fig:HSerror}, we plot the values of the left and right 
hand sides of Eq. \ref{eq:approxHS} as functions of $\Delta\tau U$ for comparison.
We consider the situations of $|n-N|=1,2$ and $3$, respectively. 
The errors of this discrete HS decomposition Eq. \ref{eq:oldHS}
depend on $|n-N|$ significantly. 
At $|n-N|=1$ and $2$, the decomposition yields values almost exact,
or, with slight deviations for $\Delta\tau U<1$.
However, at $|n-N|=3$, the deviation becomes manifest when
$\Delta\tau U>0.5$, and even more terribly, the weight becomes
negative. 

%---------------------------------------------------------------
%------------------------------------------------------------------

Therefore, we design an exact HS decomposition for the cases from SU(2)
to SU(6) in which the operator $n_i-N$ only takes eigenvalues among
$0,\pm1,\pm2$, and $\pm3$.
The form of the new HS decomposition is the same as Eq. \ref{eq:oldHS}
but it is exact.
The values of the discrete HS fields are defined as follows
\bea
\gamma(\pm 1)&=&\frac{-a(3+a^2)+d}{d}, \ \ \, \ \ \,
\gamma(\pm 2) =\frac{a(3+a^2)+d}{d}, \nn \\
\eta(\pm 1)&=&\pm\cos^{-1} \left\{ \frac{a+2a^3+a^5+(a^2-1)d}{4}
\right\} \nn\\
\eta(\pm 2)&=&\pm\cos^{-1} \left \{\frac{a+2a^3+a^5-(a^2-1)d}{4}
\right\},
\label{eq:newHS}
\eea
where $a=e^{-\frac{1}{2}\Delta\tau U}$, $d=\sqrt{8+a^2(3+a^2)^2}$.
Eq. \ref{eq:newHS} is used for all of our simulations
in 2D SU($2N$) Hubbard model in the main text.

After integrating out fermions, we arrive at the fermion
determinant whose value depends on the discrete HS fields.
The HS fields are sampled using the standard Monte Carlo technique.

%************************************************************
\section{Tests of the pinning field method}
\label{sect:test}

Below we present various tests of the pinning field method
to confirm its validity and its sensitivity to weak orderings.

%%%%%%%%%%%%%%%%%%%%%%%%%%%%%%%%%%%%%%%%%%%%%%%%%%%%%%%%%%%%%%%%%%%%%
\subsection{Test of the pinning field method in the half-filled SU(2) 
Hubbard model}
\label{sect:SU2}

%-------------------------------------------------
\begin{figure}
\includegraphics[width=\linewidth]{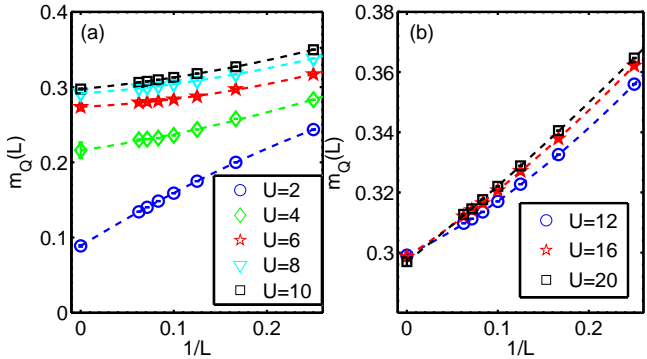}
\caption{Finite size scalings of $m_Q(L)$ v.s. $1/L$ for 
the half-filled SU(2) Hubbard model. 
The lines are fitted by the quadratic polynomial fitting
of the QMC data. 
Error bars of QMC data are smaller than symbols. }
\label{fig:neel_su2}
\end{figure}
%-------------------------------------------------

We have performed the QMC simulations with the local pinning
field method for the half-filled SU(2) Hubbard model in the
square lattice.
The finite-size scaling is presented in Fig. \ref{fig:neel_su2}.
The parameter values are the pinning field $h_{i_0j_0}=2$
and the projection time $\beta=240$.
The extrapolated values of the Neel moments $m_Q$
defined in Eq. 
increase monotonically as $U$ increases and become to 
saturate around $U=10$.
The Neel moment reaches $0.297\pm 0.002$ at $U=20$ in our simulation,
which agrees well with previous QMC simulations.
This test confirms the validity of the pinning field method.

%--------------------------------------------------------
\subsection{Sensitivity of the pinning field method to weak
ordering}
\label{sect:SU6}

%-------------------------------------------------------------
\begin{figure}[htbp]
\includegraphics[width=0.45\textwidth]{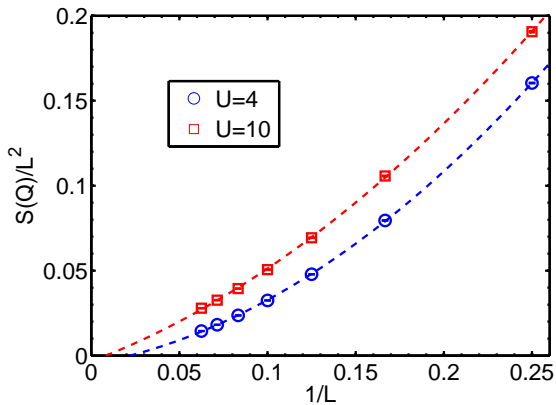}
\caption{Finite size scalings of the structure factor 
$S(Q)/L^2$ in the case of SU(6) with $U=4$ and $U=10$. 
Quadratic polynomials are used to fit the data. 
Error bars of QMC data are smaller than symbols. 
In these calculations, projection time $\beta=80$ is used. }
\label{fig:neel_ss}
\end{figure}

We consider the cases of weak Neel ordering in the half-filled SU(6) 
Hubbard model in the square lattice with $U=4$ and $U=10$.
The finite-size scalings based on structure factor are
shown in  Fig.~\ref{fig:neel_ss}.
Quadratic polynomials are used to fit the structure factor
$S(Q)/L^2$ as defined in Ref.~ \onlinecite{Cai2013a}. 
It is difficult to conclude whether long-range Neel ordering
exists or not in both cases. 
In contrast, for the case of $U=4$, 
the finite-size scaling based on the pinning field
method in Fig. 1 in the main text yields the extrapolated Neel 
moment $m_Q=0.026$.
The corresponding value of $S(Q)/L^2$ is its square at the order
of $10^{-3}$ and thus is too weak to 
identify in Fig. \ref{fig:neel_ss}.
Moreover, for the case of $U=10$ in which the largest Neel moment
appears (Fig. 4 in the main text), the corresponding structure factor
remains too small to be extrapolated through the finite size scaling.
The weak Neel orderings in the SU(6) Hubbard model were not found
in a previous work based on the structure factor method by some
of the authors either \cite{Cai2013a}. Due to the improved numeric
resolution, they are identified through the pinning field method.

%------------------------------------------------
\subsection{The pinning field method for the 
1D SU(2) and SU(4) Hubbard models}
\label{sect:1D}

%-------------------------------------------------------
\begin{figure}[htbp]
\includegraphics[width=0.5\textwidth]{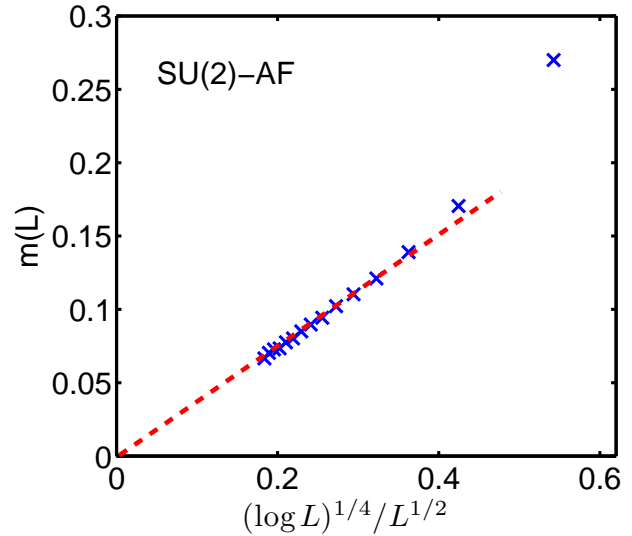}
\caption{Finite size scaling of $m(L)$ v.s. 
$(\log L)^{\frac{1}{4}}/L^{\frac{1}{2}}$ 
for the 1D half-filled SU(2) Hubbard model.
Parameter values are $\beta=80$, $U=4$ and $h_{i_0j_0}=2$.
}
\label{fig:neel_1d}
\end{figure}
%------------------------------------------------------
\begin{figure}[htbp]
\includegraphics[width=\linewidth]{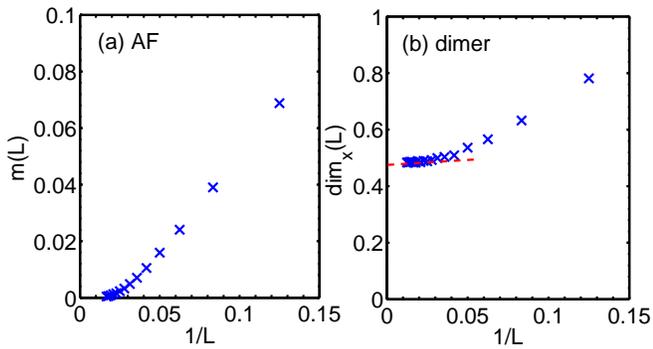}
\caption{
(a) Finite size scaling of $m(L)$ v.s. $1/L$ for the 1D half-filled SU(4)
Hubbard model.
(b) Finite size scaling of $\mbox{dim}_x(L)$ v.s. $1/L$ for the 1D 
half-filled SU(4) Hubbard model.
Parameter values are $\beta=80$, $U=4$ and $\Delta t_{i_0j_0}=2$.
}
\label{fig:su4_1d}
\end{figure}

%%%%%%%%%%%%%%%%%%%%%%%%%%%%%%%%%%%%%%%%%%

Since the pinning field method is sensitive to weak long-range
orderings, a natural question is that whether it is oversensitive.
To clarify this issue, we apply it to 1D half-filled SU(2) and SU(4)
Hubbard models in which it is well-known that magnetic
long-range orders do not exist.
The QMC simulation results presented below are in 
an excellent agreement with previous analytic and numeric results.
This confirms the validity of the pinning field method.
We use the pinning fields described in the Eq. 3 and Eq. 5 in 
the main text to investigate Neel and dimer orderings, respectively.

For the 1D half-filled SU(2) Hubbard model, the pinned sites are 
set as $i_0=1$ and $j_0=2$, respectively, and
values of the pinning fields are $h_{i_0,j_0}=2$.
We consider the induced magnetic 
moment on the furthest sites $\frac{L}{2}$ and $\frac{L}{2}+1$ 
defined as $\pm m(L)$.
Strong quantum fluctuations suppress the long-range Neel ordering, and
the asymptotic behavior of the two-point spin correlation functions
at half-filling follows the pow-law decay as
\cite{Schulz1990}
\bea
\avg{S(i) S(j)}\sim  (-)^{i-j} \frac{\log^{\frac{1}{2}} |i-j|}{|i-j|}.
\label{eq:spin1D}
\eea
Since spin moments are pinned at $i_0$ and $j_0$,
$m(L)$ should scales as 
\bea
m(L)\sim \frac{(\log L)^{\frac{1}{4}}}{\sqrt L}.
\eea
Our QMC results with pinning fields are in an excellent agreement with
Eq. \ref{eq:spin1D} as shown in Fig. \ref{fig:neel_1d}.

The magnetic properties of the 1D half-filled SU(4) Hubbard model 
are dramatically different from the SU(2) case. 
Bosonization analysis \cite{Wu2005} shows that its ground states 
exhibit long-range-ordered dimerization with a finite spin gap,
and the Neel correlation decays exponentially. 
We set the pinned sites at $i_0=1$ and $j_0=2$, respectively,
and the pinning field for dimerization as $\Delta t_{i_0j_0}=2$.
The induced dimer order is defined as the difference between
two furthest bonds $(\frac{L}{2},\frac{L}{2}+1)$ and 
$(\frac{L}{2}+1,\frac{L}{2}+2)$ as
\bea
\mbox{dim}_{i}(L) =(-)^i \big\{ d_{L/2,x}-d_{L/2+1,x}\big\}.
\eea
Our QMC simulation results are illustrated in Fig. \ref{fig:su4_1d} (b),
which exhibit the long-range ordering in agreement with previous analytic
results.

%%%%%%%%%%%%%%%%%%%%%%%%%%%%%%%%%%%%%%%%%%%%%%%%%%%%%%%%%%%
\subsection{The issue of non-linear response to the pinning field}

%-------------------------------------------------------
\begin{figure}[htbp]
\includegraphics[width=0.5\textwidth]{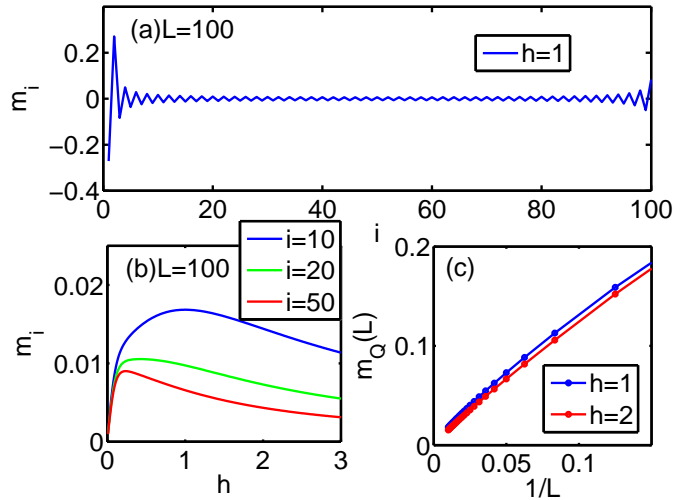}
\caption{The induced magnetic moments $m(i)$ by pinning fields in 
the non-interacting half-filled 1D SU(2) lattice model.
(a) The spacial distribution of $m(i)$ with $h_{i_0j_0}=1$
and $L=100$.
(b) The induced moments $m(i)$ v.s $h_{i_0,j_0}$ at different
sites $i=10, 20$ and $50$ in the system with $L=100$.
(c) The scaling of $m_Q(L)$ with $Q=\pi$ at two different
pinning fields. 
}
\label{fig:impurity}
\end{figure}
%------------------------------------------------------

In Fig.~1 of the main text, we present the scaling of the residual Neel 
moment $m_Q(L)$ $v.s.$ $1/L$ with two different values of the pinning fields. 
A counter-intuitive observation is that $m(L)$ is weaker at 
$h_{i_0j_0}=2$ than that of $h_{i_0j_0}=1$.
Below we present convincing evidence that actually this is not
an artifact of the finite size. 
This is a typical behavior of responses on sites far away 
from the scattering center in the strong scattering limit.

To illustrate this point, we present the calculation for a toy
model of a non-interacting half-filled SU(2) 1D lattice system,
such that we can easily calculate systems with very large size
up to $L=100$.
The pinning fields are located at sites $i_0= 1$ and $j_0=2$, 
and the induced magnetic moments $m(i)$ are presented in
Fig. \ref{fig:impurity}. 
Although it is natural that the induced magnetic moments
increase monotonically with $h$ right on the impurity sites,
there is no reason to expect the same behavior on sites away
from the scattering center.
On these sites, in fact, Fig. \ref{fig:impurity}(b) shows 
that $m(i)$'s are non-monotonic with respect to $h$.
All of them decays at large values of $h$ after passing
maxima at intermediate values of $h$. 
The finite size scalings of $m_Q(L)$ defined in the main text
are presented in Fig. \ref{fig:impurity}(c) at $h=1$ and 2.
Both curves converge to 0 as they should be in non-interacting 
systems.
Again, the curve with $h=2$ is lower than that of $h=1$.

%%%%%%%%%%%%%%%%%%%%%%%%%%%%%%%%%%%%%%%%%%%%%%%%%%%%%%%%%%%%%%%%%%%%
\section{Error analysis}
\label{sect:error}

In this section, we present the comparisons with exact diagonalization,
the analyses on errors from the discrete Suzuki-Trotter 
decomposition and finite projection time $\beta$.

\subsection{Comparison with the exact diagonalization}
%------------------------------------------------------------------
\begin{table}
\begin{ruledtabular}
\begin{tabular}{lll}
quantity & QMC & ED \\
\hline
$\langle m(1,1) \rangle_{U=4}$ & 0.4340$\pm$0.0001 & 0.4342 \\
$\langle m(3,3) \rangle_{U=4}$ & 0.2344$\pm$0.0003 & 0.2351 \\
\hline
$\langle m(1,1) \rangle_{U=12}$ & 0.4796$\pm$0.0001 & 0.4807 \\
$\langle m(3,3) \rangle_{U=12}$ & 0.3207$\pm$0.0002 & 0.3218 \\
\hline
$\langle m(1,1) \rangle_{U=20}$ & 0.4902$\pm$0.0001 & 0.4915 \\
$\langle m(3,3) \rangle_{U=20}$ & 0.3248$\pm$0.0002 & 0.3261 \\
\end{tabular}
\end{ruledtabular}
\caption{The induced magnetic moments $m(1,1)$ and $m(3,3)$ by the 
pinning fields for the half-filled SU(2) Hubbard model.
Both the QMC and exact diagonalization results are presented 
for comparison.
The parameter values are $h_{i_0j_0}=2$, $\beta=240$, $\Delta\tau=0.05$.
The lattice size is $4\times 4$. }
\label{table:ed}
\end{table}

%-------------------------------------------
\begin{figure}
\includegraphics[width=0.45\textwidth]{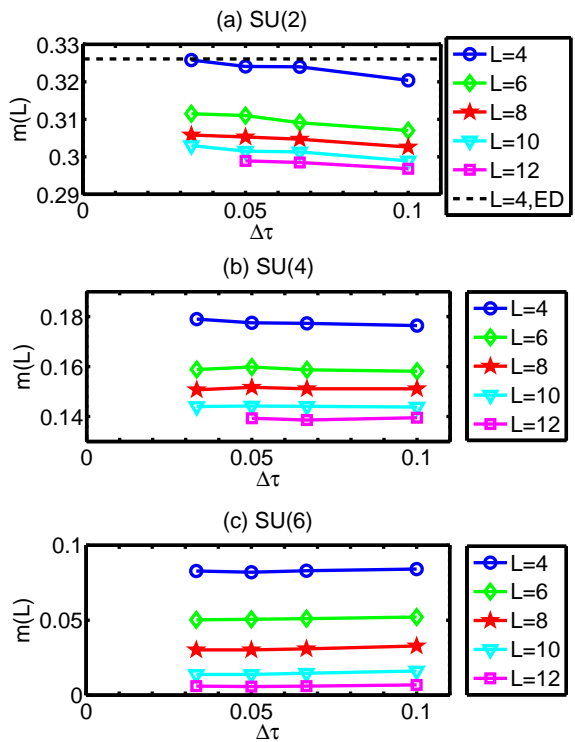}
\caption{Scaling of the Neel moments $m(L)$ v.s. $\Delta\tau$ 
for the cases of SU(2), SU(4) and SU(6) shown in 
(a)$\sim$(c), respectively. 
In the case of SU(2), exact diagonalization results are also plotted 
as the dashed line for comparison. 
The parameters are $U=20$, $\beta=80$ and $h_{i_0j_0}=2$.}
\label{fig:neel_su2n_dtau}
\end{figure}
%-------------------------------------------

%-----------------------------------------------
\begin{figure}%[htbp]
\includegraphics[width=0.5\textwidth]{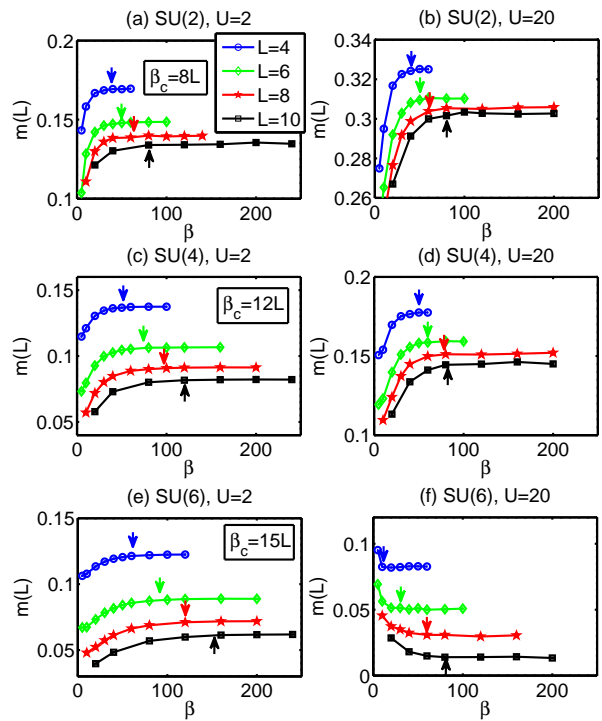}
\caption{The scalings of the Neel moments $m(L)$ v.s. $\beta$
for the half-filled SU($2N$) Hubbard model.
Lattice sizes are $L=4,6,8,10$.
The interaction parameter for (a), (c), and (e) is $U=2$, and
that for (b), (d), and (f) is $U=20$. 
Error bars of QMC data are smaller than symbols. 
The arrows mark the estimated convergence projection time $\beta_c$
of these curves.
The approximate relations of $\beta_c$ v.s $L$ are estimated as
$\beta_c=8L, 12L$ and $15 L$ for the cases of SU(2),
SU(4) and SU(6), respectively.
}
\label{fig:neel_beta}
\end{figure}
%------------------------------------------------------------

In order to check the numeric accuracy of our simulations, we 
first compare our QMC results with the pinning fields in the SU(2) 
case with those from the exact diagonalization in the $4\times4$ lattice.
\footnote{We thank H. H. Hung for providing the results of
exact diagonalization.}
The pinning fields are applied at sites $i_0=(1,1)$
and $j_0=(2,1)$ according to Eq. 3 in the main text.
In table.~\ref{table:ed}, we list the magnetic moments on sites 
$(1,1)$ and $(3,3)$ with different $U$'s.
As $U$ goes up, the numeric errors of QMC increase, 
but are still less than $0.002$ even at $U=20$.

%------------------------------------------------------
\subsection{Scaling on the discrete $\Delta \tau$}
\label{sect:errorB}

For the Suzuki-Trotter decomposition defined in Eq.~\ref{eq:trotter}, 
its error is at the order of $tU^2 (\Delta\tau)^3$.
Such an error is most severe in the large $U$ regime, and thus
we only present the scaling with respect to $\Delta\tau$ at $U=20$. 

The pinning fields are chosen in the same configuration 
described in Eq. 3 in the main text.
The distribution of $m_i$ is staggered with decaying magnitudes
as away from two pinned sites $i_0$ and $j_0$. 
The weakest moments are located at the central points  
$(\frac{L}{2}+1, \frac{L}{2}+1)$ and $(\frac{L}{2}+2, \frac{L}{2}+1)$.
The residual values at these two points are denoted as $\pm m(L)$,
respectively.
The long-range order can also be reached as the limit of $m(L)$ 
in the thermodynamic limit $L\rightarrow\infty$.

In Fig. \ref{fig:neel_su2n_dtau}, curves of
the Neel moment $m(L)$ v.s. $\Delta\tau$ are plotted 
for the three cases of SU(2), SU(4), and SU(6), respectively.
The slopes of these scaling lines are nearly independent on 
the lattice size $L$ for all three cases.
Due to convergence of the finite $\Delta\tau$ scaling, we use the 
value of $\Delta\tau=0.05$ in all our simulations.

%-----------------------------------------------------
\subsection{The finite $\beta$ scaling}

Next we check the effect of the finite projection time $\beta$.
We use the residue Neel moment $m(L)$ at the furthest points
for scaling as defined in Sect. \ref{sect:errorB}.
In Fig.~\ref{fig:neel_beta}, we present the scalings of the Neel
moments $m(L)$ v.s. $\beta$ for different sizes $L=4,6,8$, and $10$. 
For each curve, we define $\beta_c$ as the convergence projection 
time after which $m(L)$ converges, and its approximate
position is marked by an arrow.
Here we only present the scalings at $U=2$ in the weak coupling
regime and at $U=20$ in the strong coupling regime. 
The largest values of $\beta_c$ are expected in either of these two limits,
which can be understood as follows:
$\beta_c$ is determined by the finite gap of the many-body spectra.
In the small $U$ regime, the finite size gap increases as increasing
$U$, while in the large $U$ regime, it deceases as $U$ increases
because the energy scale is controlled by the magnetic exchange 
scale $J\sim 4t^2/U$.

In the case of SU(2), the relations of $\beta_c$'s on $L$ are nearly
the same for $U=2$ and $U=20$, which are estimated as $8L$. 
In the cases of SU(4) and SU(6), $\beta_c$'s at $U=2$ are larger than the 
corresponding ones at $U=20$.
At $U=2$, their dependence on $L$ is estimated as $\beta_c\approx 12L$
for the SU(4) case and $\beta_c\approx 15 L$ for the SU(6) case,
respectively.
At $U=20$, the system enters to the dimerization phase, and thus
$m(L)$ is suppressed by longer projection time.

The largest size in our simulations is $L=16$.
Considering the above scalings, we choose $\beta=15\times 16=240$ for 
all the simulations presented in the main text, which should be
sufficient to obtain accurate numeric results. 
In particular, the major result in the main text,  i.e., the non-monotonic 
behavior of $m(L=\infty)$ with increasing $U$ 
for both the SU(4) and SU(6) cases, is not 
an artifact from the finite 
projection time $\beta$.

%\bibliographystyle{prsty}
%\bibliography{sun}

\end{document}